\documentclass{article}
\usepackage{arxiv}
\usepackage[T1]{fontenc}
\usepackage[utf8]{inputenc}
\usepackage{booktabs}
\usepackage{tabularx}
\usepackage{array}
\usepackage{colortbl}
\usepackage{amsmath}
\usepackage{amssymb}
\usepackage{algorithm}
\usepackage{algorithmic}
\usepackage{xspace}
\usepackage{enumitem}
\usepackage{microtype}
\usepackage{graphicx}
\usepackage{flafter}
\usepackage{placeins}
\usepackage{float}
\usepackage[numbers,sort&compress]{natbib}
\usepackage[hidelinks]{hyperref}
\usepackage{bookmark}
\usepackage{xcolor}

\providecommand{\Description}[1]{}
\newcommand{\cineu}{\textsc{CineU}\xspace}
\newcolumntype{Y}{>{\raggedright\arraybackslash}X}
\definecolor{tablehead}{HTML}{EAF0F4}
\definecolor{tablenote}{HTML}{F5F7F9}
\newcommand{\grayrule}{\arrayrulecolor{gray}\midrule\arrayrulecolor{black}}
\definecolor{highlightcolor}{RGB}{226,232,250}

\definecolor{basecolor}{RGB}{153,153,153}

\title{Cinematize from Unreal: Agentic Choreography\\for Controllable World Simulation}
\newcommand{\CineUPDFAuthors}{Zifan Song, Caihua Shan, Kaitao Song, Yifan Yang, Dongsheng Li, Cairong Zhao}
\author{%
  Zifan Song\textsuperscript{1}\qquad
  Caihua Shan\textsuperscript{2}\qquad
  Kaitao Song\textsuperscript{2}\qquad
  Yifan Yang\textsuperscript{2}\\[0.7em]
  \bfseries Dongsheng Li\textsuperscript{2}\qquad
  Cairong Zhao\textsuperscript{1}\\[1.0em]
  \normalfont\small \textsuperscript{1}Tongji University
  \qquad \textsuperscript{2}Microsoft Research Asia
}

\date{October 2026}
\renewcommand{\shorttitle}{Cinematize from Unreal}
\hypersetup{
  pdftitle={Cinematize from Unreal: Agentic Choreography for Controllable World Simulation},
  pdfauthor={\CineUPDFAuthors},
  pdfsubject={Agentic orchestration and executable world control in Unreal Engine},
  pdfkeywords={CineU, code-based world modeling, agentic systems, executable world state, controllable world simulation, Unreal Engine}
}
\begin{document}
\maketitle
\begin{abstract}
Controllable world simulation requires more than visually plausible frames: scenes must retain persistent state, coordinated actions, and editable temporal structure. Generative world models offer flexible visual synthesis, but their outputs do not always expose the objects and transitions needed for precise revision. Graphics engines provide executable state and direct control, yet translating language into valid operations requires specialized knowledge of assets, geometry, and interdependent APIs. We present \cineu (Cinematize from Unreal), an agentic system for language-driven world orchestration and cinematic choreography within Unreal Engine. From a code-based world modeling perspective, an Unreal project serves as a persistent, inspectable world representation, while agent-generated code specifies operations whose effects can be executed and verified. CineU combines hierarchical orchestration, cascaded asset retrieval and geometry-constrained placement, capability-aligned code synthesis, and multi-stage verification with localized repair. Shared scene and timeline context coordinates entities, actions, camera trajectories, and shot boundaries while supporting dependency-aware revisions. Experiments on scene construction, selective editing, multi-shot cinematography, and component ablations show how specialized execution knowledge and engine feedback improve task completion and targeted recovery. CineU illustrates how executable world state can complement learned visual generation to support controllable and revisable simulated worlds.
\end{abstract}

\keywords{agentic systems; code-based world modeling; executable world state; controllable simulation; graphics engines.}
\section{Introduction}
\label{sec:introduction}

Building a controllable dynamic world requires more than generating visually plausible frames. A simulated scene has persistent entities, spatial relationships, actions, camera parameters, and temporal dependencies that must remain consistent as the world evolves. Graphics engines provide explicit control over these structures and support programmatic simulation~\cite{qiu2017unrealcv,ge2024behavior}. They also support cinematic previsualization, where changes to scene geometry and camera motion must be coordinated~\cite{hu2026previzwhiz}. However, translating a semantic request into a sequence of valid engine operations remains labor intensive and requires specialized knowledge.

Generative foundation models offer a complementary capability. Image, video, and interactive-world models synthesize rich visual outputs from high-level descriptions~\cite{gao2025seedream,kong2024hunyuanvideo,wan2025wan,bruce2024genie}. Graphics engines instead maintain addressable actors, asset bindings, transforms, animation tracks, and camera keyframes. These representations enable precise execution and revision, but expose a large and interdependent action space. A generated video may communicate the intended appearance of a scene without providing an executable representation of its entities and events. Conversely, an engine script may execute without realizing the intended composition. Code-based world models suggest making state and its changes explicit in executable programs, rather than relying only on visual outcomes~\cite{tang2024worldcoder,zhang2026code2worlds,chen2026codeworldbrain}. Here, the central challenge is to connect flexible semantic reasoning with controllable, verifiable operations over a persistent engine world.

\begin{figure}[t]
  \centering
  \includegraphics[width=\linewidth]{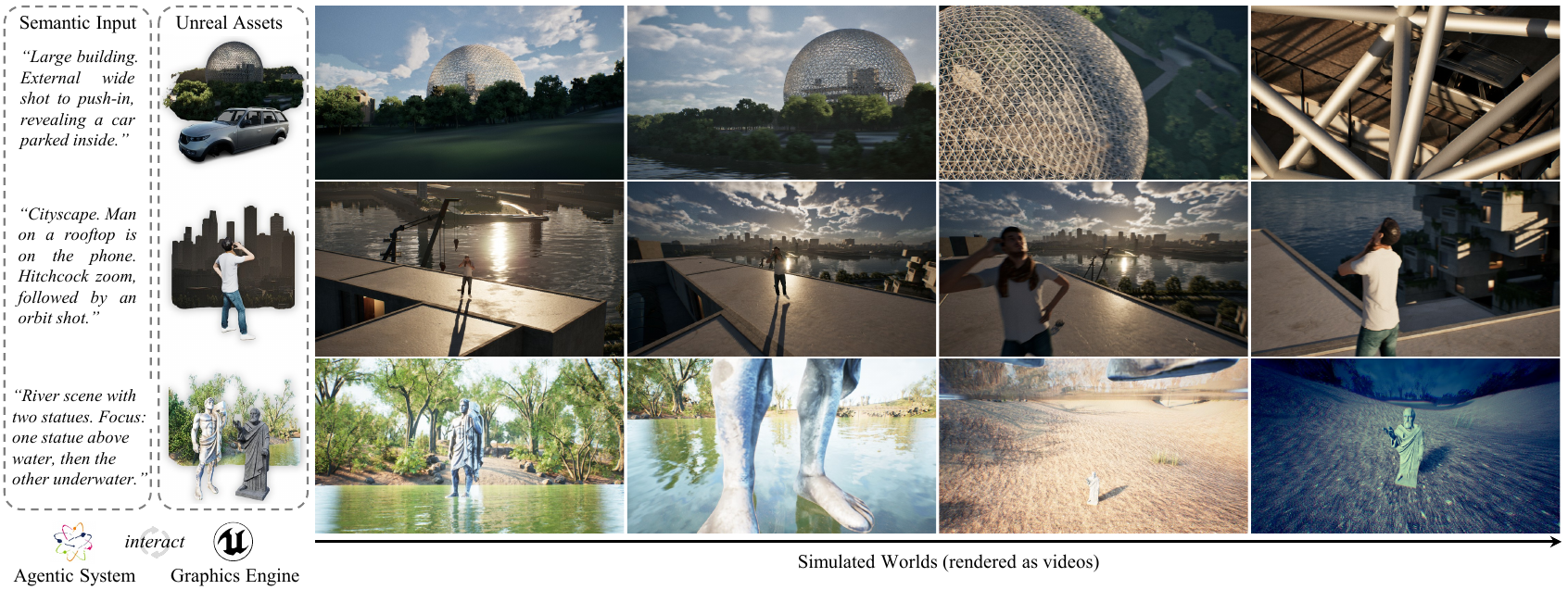}
  \caption{\textbf{Language-directed cinematic worlds.} Three briefs combine scene assets with camera and action requirements: a building reveal, a rooftop dolly zoom followed by an orbit, and attention shifting between statues above and below water. Rendered frames visualize the results; engine-native actors and tracks remain available for subsequent editing.}
  \Description{Three rows pair natural-language scene and camera briefs with asset thumbnails and rendered frames. The examples show a building and car reveal, a rooftop character with two camera movements, and statues viewed above and below water. A language-agent system communicates with the graphics engine.}
  \label{fig:overview}
\end{figure}

Recent language-agent systems combine reasoning, tool use, and feedback to execute complex tasks~\cite{yao2022react,hong2023metagpt,zhang2024aflow}. Language-guided scene systems likewise translate semantic specifications into editable environments or motion programs. UnrealLLM targets procedural scene generation in Unreal Engine~\cite{songtang2025unrealllm}; LAMP produces explicit object and camera trajectories~\cite{kizil2026lamp}; and PrevizWhiz combines rough 3D layouts with video-based previsualization~\cite{hu2026previzwhiz}. Camera-conditioned video generation further supports geometrically specified viewpoints~\cite{ren2025gen3c,zhang2026ucpe}. These approaches motivate a more general question: \emph{how can an agentic system coordinate scene construction, temporal control, and subsequent modifications within the same executable world?}

The difficulty lies in dependencies between operations. Replacing an actor may affect its support geometry, animation compatibility, and camera framing. Walking requires the actor's displacement and animation playback to agree in time. A dolly zoom couples camera distance and focal length; valid values for either parameter alone do not guarantee the desired effect. An execution error can also be localized to one operation even when previous operations have produced correct scene state. A useful agentic architecture must therefore reason about task order, shared resources, cross-track constraints, and the evidence needed to validate the resulting engine state.

We introduce \cineu (\emph{Cinematize from Unreal}), an agentic system for controllable world simulation grounded in Unreal Engine. We formulate a language instruction as a dependency graph of engine operations and execute it through specialized agents that share project and timeline context. The system supports two connected capabilities: \emph{world orchestration}, which constructs and selectively modifies environments, entities, appearances, and actions; and \emph{cinematic choreography}, which coordinates camera motion, lens control, actor animation, and multi-shot sequences. The output is a persistent Unreal project rather than an isolated rendered sequence. Unlike systems that infer world dynamics or train a new rendering model, \cineu uses generated code to control the state and capabilities of an existing simulation engine.

The architecture combines four mechanisms. First, hierarchical orchestration separates environment, entity, spatial, and Sequencer responsibilities while preserving the dependencies between them. Second, cascaded asset retrieval and geometry-constrained placement connect semantic descriptions to compatible resources and spatially feasible configurations. Third, capability-aligned code synthesis instantiates verified engine patterns using the current project's object references and timing constraints. Fourth, verification-grounded reflection checks generated operations before execution, traces their runtime behavior, inspects the resulting engine state, and repairs the responsible function where possible. Together, these mechanisms connect semantic planning, code generation, geometric reasoning, and execution feedback without assuming that a model's stated result reflects the actual simulator state.

An important consequence is that the \emph{scope of a modification depends on the relations it must preserve}. Changing a material can leave geometry and timing untouched. Replacing an entity may require a placement update to preserve support contact, while changing an animation may require re-aligning displacement and playback. The same dependency structure guides both requested modifications and error recovery. This gives the system an explicit representation of which operations can be changed independently and which must be coordinated.

We study \cineu through compound scene and camera instructions, selective revisions, multi-shot examples, and component ablations. The latter show task completion of $47.8\%$ with specialized knowledge compared with $18.4\%$ without knowledge, and $47.8\%$ with multi-stage verification compared with $16.8\%$ without verification. Function-level repair achieves $47.8\%$ completion compared with $38.1\%$ under full regeneration in the reported configurations. The qualitative examples illustrate the relationships behind these results, including camera and lens coupling, animation synchronization, and scene-preserving revisions. Section~\ref{sec:evaluation} provides the evaluation details and limitations.

Our contributions are threefold:
\begin{itemize}[leftmargin=*]
  \item \textbf{An engine-grounded agentic formulation of world simulation.} We map language instructions to dependency-aware operations over persistent scene and timeline state, supporting both construction and selective revision.
  \item \textbf{A coordinated architecture for executable choreography.} Hierarchical agents, cascaded retrieval, spatial constraints, capability-aligned code generation, and verification-guided local repair connect specialized operations through shared execution context.
  \item \textbf{An empirical analysis of control and recoverability.} Cinematic examples and component ablations characterize cross-track synchronization, revision dependencies, the effect of specialized knowledge, and the contribution of verification and targeted repair.
\end{itemize}

Rather than replacing a graphics engine with a learned simulator, \cineu uses agentic reasoning to access and coordinate its explicit controls. This combination offers a practical route from high-level instructions to worlds whose scene state, timing, and execution outcomes remain inspectable and revisable.

\section{Related Work}
\label{sec:related}

\subsection{Agentic Orchestration and Specialized Execution}

Language-based reasoning and agentic execution provide complementary building blocks. Chain-of-thought prompting and Graph of Thoughts organize reasoning steps~\cite{wei2022COT,besta2024graph}, while ReAct interleaves reasoning with environmental actions~\cite{yao2022react}. Mindstorms explores natural-language societies of interacting modules~\cite{zhuge2023mindstorms}. Multi-agent frameworks develop these ideas through role-playing communication in CAMEL~\cite{li2023camel}, expert collaboration in AgentVerse~\cite{chen2024agentverse}, conversational composition in AutoGen~\cite{wu2024autogen}, and structured software-development processes in MetaGPT and ChatDev~\cite{hong2023metagpt,qian2024chatdev}. Work on scaling collaboration and Graph-of-Agents further examines how communication structure and agent selection shape joint problem solving~\cite{qian2024scaling,yun2026graphagents}.

Automation can also operate on the workflow itself. Promptbreeder and optimization by language models modify prompts~\cite{chirs2024probreed,cheng2024opro}; GPTSwarm represents language agents as optimizable graphs~\cite{zhuge2024gptswarm}. Automated Design of Agentic Systems and AFlow search over agent implementations and workflows~\cite{hu2024automated,zhang2024aflow}, while Archon searches compositions of inference-time techniques~\cite{saad2024archon}. The Workforce architecture separates planning, coordination, and tool-using workers; OWL trains its planner through reinforcement learning~\cite{hu2025owl}. These approaches motivate explicit orchestration, but optimizing a task-solving workflow differs from preserving validated scene state while applying a constrained update.

Domain-specific systems illustrate why execution context matters. CogAgent addresses visual GUI operation~\cite{hong2024cogagent}; MobileExperts and Mobile-Agent-v2 organize mobile-device operations through specialized agents~\cite{jia2024mobile,wang2024mobile}; and web agents with world models use environment dynamics for navigation~\cite{chaeweb}. Data Interpreter, natural-language-to-SQL methods, and Trade in Minutes target data analysis, database access, and financial decision workflows, respectively~\cite{hong2024data,bo2024nlsql,song2025timi}, while Z-AXIS applies agentic reasoning to enterprise evaluation~\cite{song2026zaxis}. In mathematical tasks, problem-understanding methods and DeepSeek-Prover specialize reasoning for word problems and formal proofs~\cite{zhong2024achieving,xin2024deepseekprover}. Their shared relevance is the need to connect language to domain-specific objects, tools, and validity conditions; their task outcomes are distinct from cinematic authoring.

Code capability is another separate concern. AlchemistCoder develops code-model capability through hindsight tuning~\cite{song2024alchemistcoder}, whereas LiveCodeBench evaluates code generation through executable tasks~\cite{jain2024livecodebench}. In \cineu, a generated program is an intermediate operation over a persistent scene. Correct API use must therefore be coordinated with asset bindings, spatial relations, and shared timeline state. Our hierarchy follows these editable dimensions, and its shared context connects operations whose outputs must agree.

\subsection{Memory, Retrieval, and Verification-Guided Repair}

Persistent context and retrieved knowledge serve different roles. InfLLM addresses efficient access to long-context information~\cite{xiao2024infllm}; G-Memory organizes cross-trial collaboration histories for multi-agent systems~\cite{zhang2025gmemory}. MindSearch distributes information seeking and integration across agents~\cite{chen2025mindsearch}. \cineu separates a library of reusable engine patterns from the current project's realized state: a valid code pattern does not identify which actor, binding, or timeline a new operation must modify.

Retrieval itself does not establish correctness. Self-RAG learns to retrieve on demand and uses reflection tokens to assess retrieved passages and generated output~\cite{asai2024selfrag}. In an engine-grounded task, the corresponding distinction is between retrieving a relevant template and verifying its instantiated effect. The former supplies a candidate operation; the latter requires execution feedback and inspection of the resulting world.

Self-Refine and language-model self-debugging use feedback to revise an initial output~\cite{madaan2023self,chen2023teaching}. Debug like a Human emphasizes stepwise runtime verification for program repair~\cite{li2024debug}. Analyses of verification-based test-time computation and faulty-agent collaboration make feedback quality and error propagation explicit research concerns~\cite{setlur2025scaling,huang2025resilience}. \cineu applies staged checks to generated engine operations and directs revision toward the failing function. The resulting repair boundary is consequential: repairing a camera operation should retain valid scene construction whenever the affected operation can be isolated.

\subsection{World Models and Executable Simulation}

World simulation encompasses several representations with different uses. Dreamer and its subsequent world-model developments learn dynamics for behavior learning~\cite{hafner2019dream,hafner2023mastering}. Learning Interactive Real-World Simulators and Genie generate interactive environments from learned visual priors~\cite{yang2023learning,bruce2024genie}. Diffusion-based game engines and DIAMOND model visual dynamics~\cite{valevski2024diffusion,alonso2024diffusion}; GEM adds controls over ego-motion, object dynamics, and scene composition~\cite{hassan2025gem}. DINO-WM uses pretrained visual features for planning~\cite{zhou2024dino}, and exploration-driven generative environments examine interaction with generated worlds~\cite{savov2025exploration}. These systems relate visual state to actions, rather than treating simulation as rendering alone.

Code-based world modeling makes state changes explicit through executable representations. WorldCoder writes programs to model an environment from interaction~\cite{tang2024worldcoder}; Code World Models for General Game Playing synthesizes transition rules that can be used by planners~\cite{lehrach2026codeworldmodels}. Code2Worlds generates simulation code for dynamic 4D scenes and refines it through physics-aware feedback~\cite{zhang2026code2worlds}. The Code World Model framework separates persistent state evolution, maintained by a coding agent, from video-based visual realization~\cite{chen2026codeworldbrain}, while Code-as-World discovers executable physical representations from multimodal observations~\cite{wang2026codeasworld}. These approaches share a commitment to executable structure but differ in how dynamics are obtained and how worlds are rendered or controlled.

Engineered platforms expose a complementary form of control. UnrealCV, PUG, BEHAVIOR Vision Suite, and Unreal-MAP support programmatic observation, controllable synthetic data, or multi-agent simulation~\cite{qiu2017unrealcv,bordes2023pug,ge2024behavior,hu2025unreal}. BEHAVIOR-1K supplies a broad embodied-activity setting~\cite{li2023behavior}. Application-specific work includes synthetic person re-identification in UnrealPerson~\cite{zhang2021unrealperson}, closed-loop sensor simulation in UniSim~\cite{yang2023unisim}, and generative autonomous-driving modeling in GenAD~\cite{zheng2024genad}. Dataset customization and perturbation offer further approaches to visual-data construction~\cite{song2024diverse,song2025dpl++}. These applications motivate explicit scene and action control, while \cineu focuses on the executable operations that construct and revise such scenes.

The distinction between learned and executable worlds is not binary. Video2Game and PhysGen3D reconstruct interactive environments from visual input~\cite{xia2024video2game,chen2025physgen3d}. \cineu does not infer new transition rules or train a visual world model. Instead, it uses code to orchestrate the existing abstractions of a graphics engine, retaining an authored scene and timeline. Its focus is how language-driven operations compose, remain revisable, and recover from execution failures in that persistent representation.

\subsection{Language-Guided Scene Construction and Editing}

Language-driven scene construction already connects semantic specifications to structured environments. Holodeck generates embodied 3D environments using language guidance~\cite{yang2024holodeck}, and ChatScene uses domain knowledge for safety-critical driving scenarios~\cite{zhang2024chatscene}. UnrealLLM is especially close to our implementation setting, combining language agents, domain knowledge, and procedural content generation in Unreal with editable engine representations~\cite{songtang2025unrealllm}. These systems establish important precedents for language-mediated access to simulation and scene-building tools.

Recent agentic methods further integrate spatial and visual feedback. Scenethesis combines language and vision for scene generation~\cite{ling2026scenethesis}; spatially contextualized VLMs maintain geometric and relational context for generation and downstream editing~\cite{liu2026agentic3d}. SAGE couples scene generators with critics for task-conditioned, simulation-ready environments~\cite{xia2026sage}. Edit-As-Act uses goal-regressive planning and geometric validation for selective indoor-scene editing~\cite{noh2026editasact}. \cineu builds on this broader direction but concentrates on the coupling between scene changes and animation, transform, lens, and cut tracks. An entity replacement is therefore evaluated in the context of the authored action and shot, not only the static layout.

\subsection{Controllable Visual Generation and Cinematography}

Controllable synthesis offers a range of ways to express visual intent. GLIGEN grounds image generation in spatial conditions~\cite{li2023gligen}, Composer combines multiple conditions~\cite{huang2023composer}, and ControlNet adds structural conditioning to diffusion models~\cite{zhang2023adding}. DreamBooth supports subject-driven generation~\cite{ruiz2023dreambooth}, while InstructPix2Pix expresses edits through language~\cite{brooks2023instructpix2pix}. Seedream develops general image synthesis~\cite{gao2025seedream}, and Generative Photography studies scene-consistent camera control~\cite{yuan2025generative}. These approaches distinguish controls over identity, appearance, layout, and viewpoint, dimensions that also need distinct revision targets in a 3D authoring tool.

Video systems add temporal dependencies. Caption-based video summarization also connects temporal media to language~\cite{zhao2024cap2sum}. AnimateDiff animates personalized image-generation models~\cite{guo2023animatediff}, and VideoComposer combines motion and other conditioning signals~\cite{wang2023videocomposer}. Snap Video develops spatiotemporal modeling~\cite{menapace2024snap}; Collaborative Video Diffusion addresses consistency across camera-controlled videos~\cite{kuang2024collaborative}; and StreamingT2V extends text-conditioned video generation over time~\cite{henschel2025streamingt2v}. HunyuanVideo, Wan, Seedance, and Veo provide general text-conditioned video synthesis~\cite{kong2024hunyuanvideo,wan2025wan,gao2025seedance,veo3}. Analysis from a physical-law perspective highlights that visual plausibility and physical consistency are separate properties~\cite{kang2024far}.

Camera and geometric control extend beyond prompts. AC3D analyzes camera control in video diffusion transformers~\cite{bahmani2025ac3d}; Gen3C supplies 3D-informed guidance~\cite{ren2025gen3c}; and Generative Rendering uses a 4D representation to guide video synthesis~\cite{cai2024generative}. Unified Camera Positional Encoding and 3D Scene Prompting encode explicit camera or scene conditions~\cite{zhang2026ucpe,lee2026sceneprompt}. Infinity-RoPE supports action changes and cinematic cuts during long video rollout~\cite{yesiltepe2026infinityrope}, while DragStream enables streaming drag-based manipulation~\cite{zhou2026dragstream}.

LAMP is particularly relevant: it translates language into a motion domain-specific language and deterministically derives object and camera trajectories that guide video generation~\cite{kizil2026lamp}. Building on this connection between language and explicit motion, \cineu focuses on the persistent engine project in which assets, placement, animation bindings, lens parameters, and shot boundaries are jointly represented, together with verification and function-level repair of operations over that project. PrevizWhiz further demonstrates the value of combining rough 3D scenes and 2D video with generative previsualization~\cite{hu2026previzwhiz}. The design question is consequently which representation supports the creator's next decision: a generated preview can guide appearance, while an engine-native scene and timeline provide concrete objects for continued revision.

\subsection{Structured Editing and User Control}

Interactive systems provide useful precedents for keeping generated results interpretable and revisable. Human--AI interaction guidelines emphasize clear system behavior and correction~\cite{amershi2019guidelines,yang2020reexamining}; mixed-initiative interaction and the COFI framework examine how automated contributions are combined with user decisions~\cite{horvitz1999principles,rezwana2023cofi}. In language and visualization systems, Wordcraft, AI Chains, and HAIChart expose intermediate steps or alternative outputs that can be inspected and revised~\cite{yuan2022wordcraft,wu2022aichains,yu2024hai}. Studies of control across co-creation phases further suggest that the desired scope of intervention changes as a result becomes more complete~\cite{issak2026control}.

Spatial authoring systems provide a closer comparison. VRCopilot uses intermediate wireframes for scene layout~\cite{zhang2024vrcopilot}; Dreamcrafter supports 3D editing through proxy representations~\cite{vachha2024dreamcrafter}; and ImaginateAR combines generated assets with scene understanding and manual refinement~\cite{lee2025imaginatear}. Generative Agents instead simulates character behavior through memory, planning, and reflection~\cite{park2023generative}. These works motivate preserving addressable structure after generation. Our focus is on the execution dependencies that make such structure usable in a graphics engine, particularly the coupling of actors, transforms, animation tracks, and camera timing.

\section{System Requirements}
\label{sec:design}

A language instruction describes objects and relationships, whereas an executable Unreal operation requires compatible assets, object references, transforms, bindings, and frame ranges. This gap gives rise to four requirements for a controllable simulation system. Table~\ref{tab:design-goals} connects each requirement to its implementation and the available evidence. The requirements follow from the dependency structure of engine tasks and the observed failure modes, including invalid placement, inconsistent tracks, and unsuccessful repair.

\begin{table}[!htbp]
\caption{\textbf{System requirements and corresponding mechanisms.} The requirements concern translating semantic instructions into operations, maintaining existing state, diagnosing failures, and coordinating coupled controls.}
\Description{Four rows map task decomposition, selective updates, diagnostic validation, and synchronization to implementation choices and relevant experiments.}
\label{tab:design-goals}
\begingroup
\small
\renewcommand{\arraystretch}{1.14}
\setlength{\tabcolsep}{5pt}
\begin{tabularx}{\linewidth}{>{\raggedright\arraybackslash}p{0.19\linewidth}YY}
\toprule
\rowcolor{tablehead}\textbf{Requirement} & \textbf{Implementation} & \textbf{Evidence} \\
\midrule
\textbf{R1: Ground instructions} & Decompose a brief into operations and dependencies; retain engine objects and runtime context. & Compound scene and camera tasks (Section~\ref{sec:expressiveness}). \\
\textbf{R2: Preserve valid state} & Update targeted materials, bindings, or tracks while retaining unrelated operations. & Texture, animation, and entity revisions (Section~\ref{sec:selective-revision}). \\
\textbf{R3: Localize failures} & Check code, trace runtime calls, inspect engine state, and repair the affected function. & Verification and repair ablations (Table~\ref{tab:ablation}). \\
\textbf{R4: Synchronize controls} & Share timeline references and constraints across animation, transform, and camera operations. & Multi-shot, dolly-zoom, and locomotion cases (Sections~\ref{sec:expressiveness} and~\ref{sec:coordination}). \\
\bottomrule
\end{tabularx}
\endgroup
\end{table}

\paragraph{R1: Ground instructions in engine state.}
A language specification is not an execution record. Task decomposition must identify which entities and operations a request requires, while engine queries determine the currently realized objects, available resources, and timeline state. The system must distinguish intended behavior from what has actually executed.

\paragraph{R2: Preserve valid state during revision.}
An operation should update only the state necessary to satisfy the new requirement. A material substitution need not regenerate animation or camera tracks. An entity substitution, however, can change support geometry and animation compatibility. Preservation therefore means retaining unaffected constraints and verified operations, not keeping every numerical value unchanged.

\paragraph{R3: Convert failures into localized feedback.}
Syntax errors, invalid API calls, missing assets, incorrect bindings, and desynchronized tracks require different corrections. A recovery mechanism needs evidence from static checks, execution traces, and the resulting engine state. Bounded function-level revision and rollback can avoid reopening validated parts of the task.

\paragraph{R4: Coordinate dependent temporal controls.}
Animation and displacement must refer to compatible motion and time intervals. A dolly zoom requires coupled camera distance and focal length. The Sequencer timeline supplies shared references for these operations, allowing specialized agents to modify individual tracks without losing their joint constraints. Reproducibility depends on the committed project state and configuration, while verification checks whether generated operations satisfy the intended constraints.

\section{Task Representation and Execution Workflow}
\label{sec:interaction}

\cineu converts a natural-language specification into an ordered set of operations over a persistent Unreal project. Its workflow maintains separate representations for the requested task, the realized engine state, and the execution context that connects them. This separation supports incremental changes, cross-track coordination, and diagnosis when an operation does not produce the intended result. Section~\ref{sec:system} describes the corresponding agents and middleware; Section~\ref{sec:evaluation} examines their behavior in scene and camera tasks.

\subsection{Task, World State, and Execution Context}

The \emph{task specification} identifies entities, spatial relationships, actions, camera behavior, and relevant constraints. The \emph{world state} consists of realized actors, transforms, assets, material assignments, and Sequencer tracks. The \emph{execution context} holds task dependencies, engine object references, timeline identifiers, and diagnostic feedback. These representations are related but not interchangeable: an accepted instruction does not imply that the corresponding operation has completed in the engine. In this executable-world view, agent-generated code proposes an intervention, while Unreal performs the state transition. The stored project state, rather than the agent's textual prediction, is the basis for subsequent operations.

The operation types mirror the engine's controllable units. Environment tasks configure levels and atmosphere; entity tasks retrieve and import resources; spatial tasks satisfy placement constraints; action tasks bind animation and displacement; and camera tasks specify framing, lens values, trajectories, and cuts. Each type has specialized implementation requirements. Their inputs and outputs are nevertheless coupled, since an animation requires a compatible actor and a camera trajectory depends on the subject's actual position.

\subsection{Execution Sequence}

\paragraph{Decomposition and grounding.}
An orchestration agent maps the instruction to a dependency graph of environment, entity, placement, action, and camera subtasks. Asset retrieval first uses the validated local cache and then searches external resources if necessary. Middleware queries the active project for object references, support geometry, and timeline state. These queries replace implicit assumptions about the scene with concrete engine context.

\paragraph{Synthesis and execution.}
Specialized agents instantiate capability-specific code patterns. The Sequencer agent establishes a shared timeline, and camera, animation, and transform sub-agents operate on its track references. Middleware performs static checks, executes the generated functions, and collects runtime traces and engine-state feedback. A successfully executed function can still violate a spatial or temporal requirement, so validation extends beyond syntax and runtime completion.

\paragraph{State inspection and revision.}
The resulting actors, keyframes, and camera tracks persist in Unreal. Subsequent instructions can target a material assignment, entity binding, or track without reconstructing the entire world. The viewport shows the rendered state, while Sequencer and execution traces expose the structures and operations relevant to validation or recovery.

\subsection{Local Updates and Dependency Scope}

A modification has an explicit \emph{target} and an associated \emph{dependency scope}. Selective invocation addresses the target resource; shared execution context identifies other state that must be reconsidered. An animation replacement may require displacement alignment, and an entity replacement may require updated support constraints. By contrast, a material-only change can usually leave geometry and timing unchanged. The system therefore aims to preserve valid relationships and unrelated operations rather than enforce strict parameter-level isolation.

\begin{table}[!htbp]
\caption{\textbf{Revision operations and dependency scope.} The addressed resource identifies where an edit starts; the retained relationship determines which surrounding state must be reconsidered. A single instruction can require changes to multiple dependent resources.}
\Description{Five rows distinguish revision targets, editable engine resources, relationships to retain, and dependent state to reconsider. Material edits retain geometry and timing. Animation edits require motion and displacement to agree. Entity replacement requires support and animation compatibility. Dolly-zoom camera edits couple distance and focal length. Shot-interval edits connect adjacent cuts to action timing.}
\label{tab:revision-dependencies}
\begingroup
\small
\renewcommand{\arraystretch}{1.16}
\setlength{\tabcolsep}{5pt}
\begin{tabularx}{\linewidth}{>{\raggedright\arraybackslash}p{0.15\linewidth}YYY}
\toprule
\rowcolor{tablehead}\textbf{Revision target} & \textbf{Addressable resource} & \textbf{Relationship to retain} & \textbf{Dependent state to reconsider} \\
\midrule
\textbf{Material} & Material assignment or appearance parameters. & The same actor remains in the same scene and temporal sequence. & Retain geometry, transform, animation, and camera tracks when only appearance changes. \\
\addlinespace[3pt]
\textbf{Animation} & Animation binding and its track section. & Pose progression agrees with the actor's displacement and action interval. & Align clip playback with the existing transform track and shared timing. \\
\addlinespace[3pt]
\textbf{Entity} & Actor or asset binding and placement transform. & The replacement retains the intended support and scene relationship. & Revisit bounds, support contact, and compatibility with the assigned animation. \\
\addlinespace[3pt]
\textbf{Camera} & Camera transform and lens channels. & In a dolly zoom, subject framing follows the distance--focal-length relation. & Update the coupled tracks together, using the subject position and common time reference. \\
\addlinespace[3pt]
\textbf{Shot interval} & Camera-cut segment and associated interval. & Shot ordering remains coherent with the action being shown. & Reconcile adjacent cut boundaries with the shared timeline and action timing. \\
\bottomrule
\end{tabularx}
\endgroup
\end{table}

Table~\ref{tab:revision-dependencies} summarizes the dependency patterns encountered in the evaluated revision tasks. These relations guide which agent is invoked, what context it receives, and which post-execution checks are required. In particular, a local edit may legitimately change multiple linked tracks if that is necessary to maintain the intended event.

\begin{figure}[t]
  \centering
  \includegraphics[width=\linewidth]{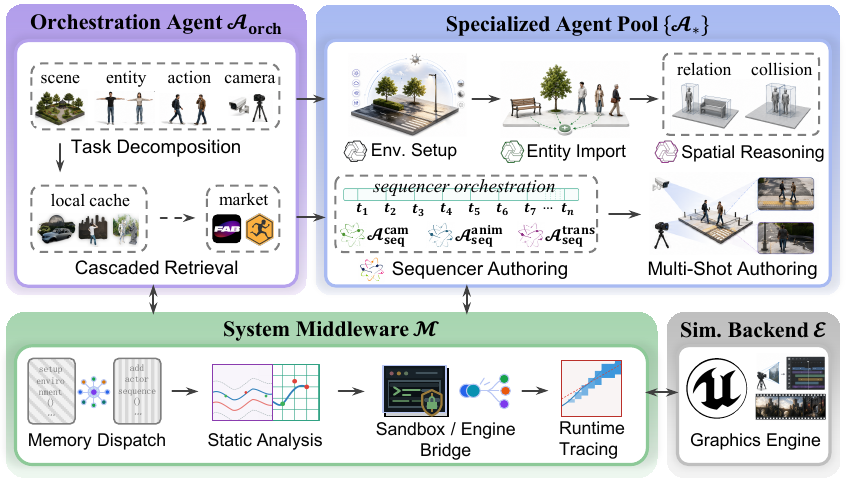}
  \caption{\textbf{Multi-tier architecture of \textsc{CineU}}, comprising specialized agents, system middleware, and simulation backend. The orchestration agent decomposes briefs and manages cascaded asset retrieval. Specialized agents configure environments, import entities, reason about placement, and coordinate Sequencer tracks. Middleware dispatches capability-specific knowledge, checks code, bridges execution to Unreal, and records runtime traces. Engine-state inspection closes the verification and repair loop.}
  \Description{The multi-layer architecture has four blocks. Orchestration at upper left includes scene, entity, action, and camera requirements, task decomposition, a local cache, and a marketplace. The specialized pool at upper right includes environment, entity, spatial, and Sequencer agents, with camera, animation, and transform sub-agents. Middleware below handles memory dispatch and isolation, static checks, sandboxed engine bridging, and runtime tracing. A separate Unreal backend exchanges state and operations with middleware.}
  \label{fig:architecture}
\end{figure}

\subsection{Temporal Composition and Multi-Shot Execution}

A shot is represented by a camera actor, a time interval, and a camera-cut track segment. All shots share a common Sequencer timeline. Shot transitions can therefore be aligned with entity motion and action changes instead of being assembled as unrelated clips. Changing a shot interval may require updates to adjacent cuts and action timing. This dependency is explicit in the engine representation and can be checked after execution. Section~\ref{sec:expressiveness} examines a multi-shot case.

\subsection{Example: A Rooftop Sequence}

The rooftop instruction in Figure~\ref{fig:overview} combines a rooftop character, an action, a dolly zoom, and an orbit. The task graph orders scene construction, entity retrieval, placement, animation binding, and camera segments. The spatial agent derives placement from the building and actor bounds; Sequencer supplies a shared time reference for the action and the two camera movements. The dolly zoom couples focal length with camera distance, while the orbit changes the viewpoint around the subject.

This example illustrates why correct isolated operations are insufficient. Camera parameters depend on the realized actor position, shot segments must agree with sequence timing, and a failed camera function should not invalidate successful environment construction. The same stored dependencies support targeted modifications and function-level repair, with verification applied to the affected state.

\FloatBarrier

\section{System Design and Implementation}
\label{sec:system}

\subsection{Architecture and Responsibility Boundaries}

\cineu separates orchestration, specialized generation, middleware, and engine execution (Figure~\ref{fig:architecture}). The architecture follows the task and state representations in Section~\ref{sec:interaction}. It assigns operations to appropriate specialized capabilities while maintaining shared world state and timeline references.

\paragraph{Task model.}
We model the execution setting as a tuple $(\mathcal{T},\mathcal{E},\mathcal{A},\mathcal{O},\mathcal{W})$. Here, $\mathcal{T}$ is the natural-language task specification encoding scene, action, and camera intent; $\mathcal{E}$ is the graphics-engine environment; $\mathcal{A}$ is the agent pool; $\mathcal{O}$ is the set of atomic engine operations exposed through the system; and $\mathcal{W}$ is the realized world state, including actors, transforms, animation bindings, camera parameters, and Sequencer tracks.

\paragraph{Orchestration agent.}
The orchestration agent interprets the task specification, decomposes it into subtasks, identifies precedence relationships, and manages asset retrieval. Its dependency graph determines the order and context of specialized invocations. A camera operation that follows a subject depends on the corresponding entity and placement operations; animation and transform operations depend on the shared sequence initialization.

\paragraph{Specialized agent pool.}
The environment agent configures the level and atmosphere. The entity agent imports and integrates assets. The spatial agent determines transforms that satisfy relational and collision constraints. The Sequencer agent initializes the timeline and coordinates camera, animation, and transform sub-agents. The separation limits irrelevant task context, while parent-provided references and synchronization constraints connect work that must remain coupled.

\paragraph{System middleware and backend.}
Middleware provides file and process operations, backend interaction, code validation, tracing, and two-version reflection rollback. Agent instructions use XML-formatted messages; parameters and serialized state use JSON schemas. Unreal Engine, accessed through its Blueprint system and programmatic interfaces, contains the realized actors, assets, and timeline. The implementation uses Unreal Engine 5.5 and earlier versions described in the system configuration. We rely on engine queries for realized state rather than treating a model's textual description as evidence that an operation succeeded.

\paragraph{End-to-end execution chain.}
We summarize the system-level transformation from an authoring request to a realized engine state as
\begin{equation}
\Phi: \mathcal{T}
\xrightarrow{\mathcal{A}_{\mathrm{orch}}}
\mathcal{G}
\xrightarrow{\{\mathcal{A}_*\}}
\{\mathcal{O}_*\}
\xrightarrow{\mathcal{M}}
\mathcal{E}(\mathcal{O}_*)
\rightarrow \mathcal{W}.
\label{eq:execution_chain}
\end{equation}
The orchestration agent $\mathcal{A}_{\mathrm{orch}}$ decomposes $\mathcal{T}$ into a dependency graph $\mathcal{G}$. The specialized set $\{\mathcal{A}_*\}$ includes the environment, entity, spatial, and Sequencer agents,
\[
\{\mathcal{A}_*\}=\{\mathcal{A}_{\mathrm{env}},\mathcal{A}_{\mathrm{entity}},\mathcal{A}_{\mathrm{spatial}},\mathcal{A}_{\mathrm{seq}},\ldots\},
\]
and maps graph subtasks to engine operations $\{\mathcal{O}_*\}\subseteq\mathcal{O}$. Middleware $\mathcal{M}$ validates, executes, and inspects these operations through $\mathcal{E}$, producing the resulting world state $\mathcal{W}$. This chain connects the architecture in Figure~\ref{fig:architecture} to the dependency, placement, synthesis, and verification mechanisms below. It can be understood as code-mediated world control: the agents generate operations, while the engine provides the state transition and its observable result. The code is therefore a control interface to an executable world, not a separately learned model of its dynamics.

\subsection{Task Dependencies and Execution Context}

The dependency graph $\mathcal{G}=(\mathcal{V},\mathcal{D})$ contains subtasks and precedence relationships. Typical subtasks include environment configuration, entity import, placement, animation binding, transform animation, camera construction, and shot coordination. Each subtask corresponds to a bounded set of engine operations. Their dependencies identify both ordering requirements and information that one operation must supply to another.

Middleware retrieves project state and environmental attributes from Unreal to contextualize those subtasks. During temporal authoring, the Sequencer agent provides a timeline reference, duration, and track allocation to its sub-agents. The generated camera, transform, and animation functions inherit this context rather than independently create incompatible sequence structures. State inspection after execution provides the next repair attempt with concrete diagnostic information.

Selective invocation is grounded in subtask responsibilities and the current engine context. Cross-track relationships are maintained through shared references and synchronization constraints; after a change, the affected operations are checked against the requested scene and timeline relationships.

\begin{figure}[t]
  \centering
  \includegraphics[width=0.93\linewidth]{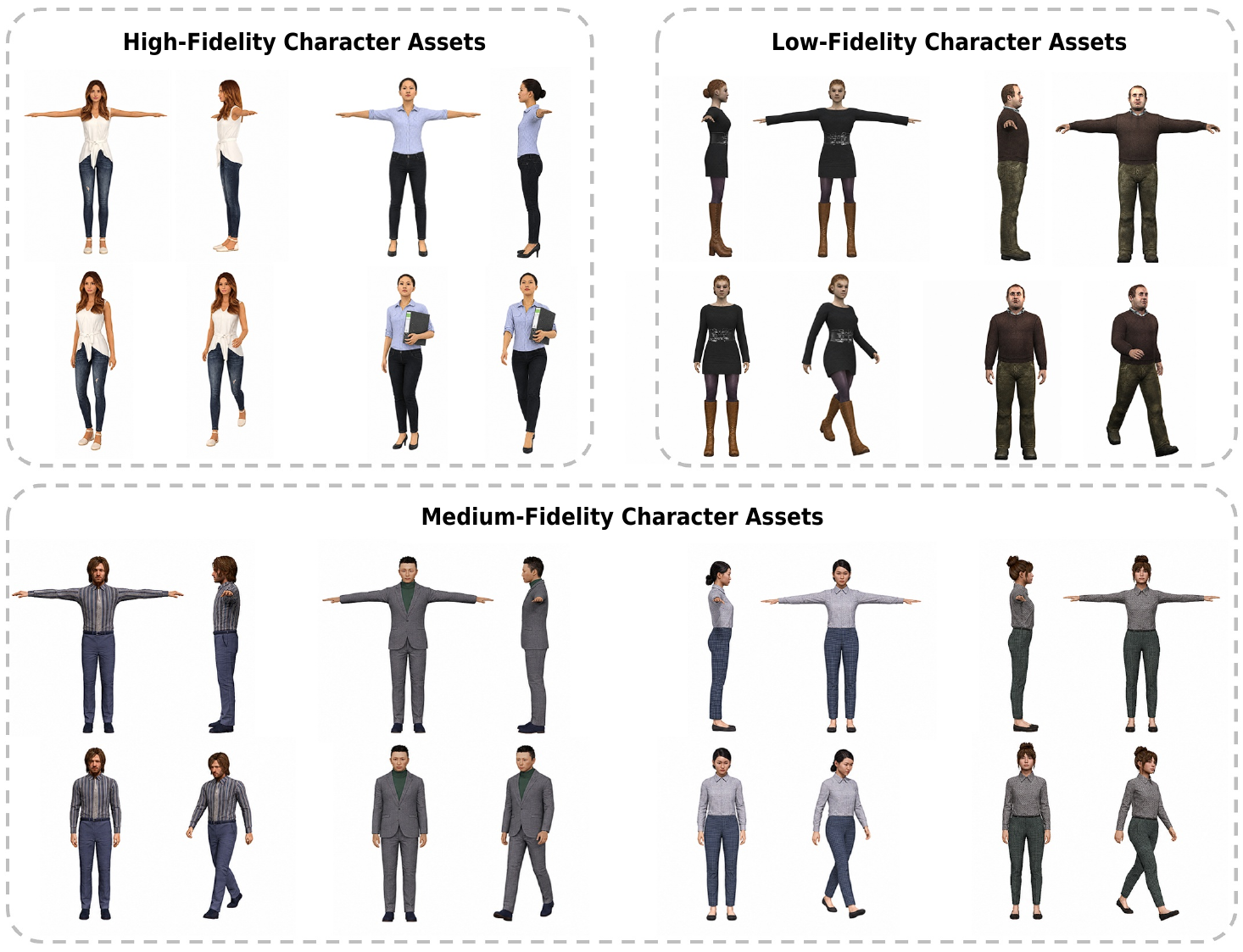}
  \caption{\textbf{Character resources at different visual-fidelity levels.} High-, medium-, and low-fidelity groups show character resources in front and side views and in representative poses. Appearance selection provides the entity resource used by subsequent placement and animation operations.}
  \Description{Three dashed groups contain human character models at high, low, and medium visual-fidelity levels. Each group includes front and side views and standing or walking poses.}
  \label{fig:character-fidelity}
\end{figure}

\subsection{Cascaded Asset Retrieval}

The local cache contains more than 300 validated assets, primarily sourced from Unreal Fab and Mixamo and annotated with semantic tags and functional descriptors. Retrieval first matches the requested asset against this cache. When no suitable local resource is found, marketplace search is invoked and refined using the engine version and project constraints. Validated external assets are cached for reuse. The cache provides known compatible resources; marketplace retrieval expands the set of realizable scene entities.

An imported resource remains an asset binding in the scene rather than becoming inseparable from a rendered output. This makes replacement possible without rebuilding all surrounding content, but also makes compatibility consequential: a replacement can change geometry, support relationships, or the animation requirements of the target entity. Asset availability constrains the realizable scene: execution depends on retrievable geometry, compatible animations, and supported engine operations.

\subsection{Character Appearance and Action Resources}

\begin{figure}[t]
  \centering
  \includegraphics[width=0.9\linewidth]{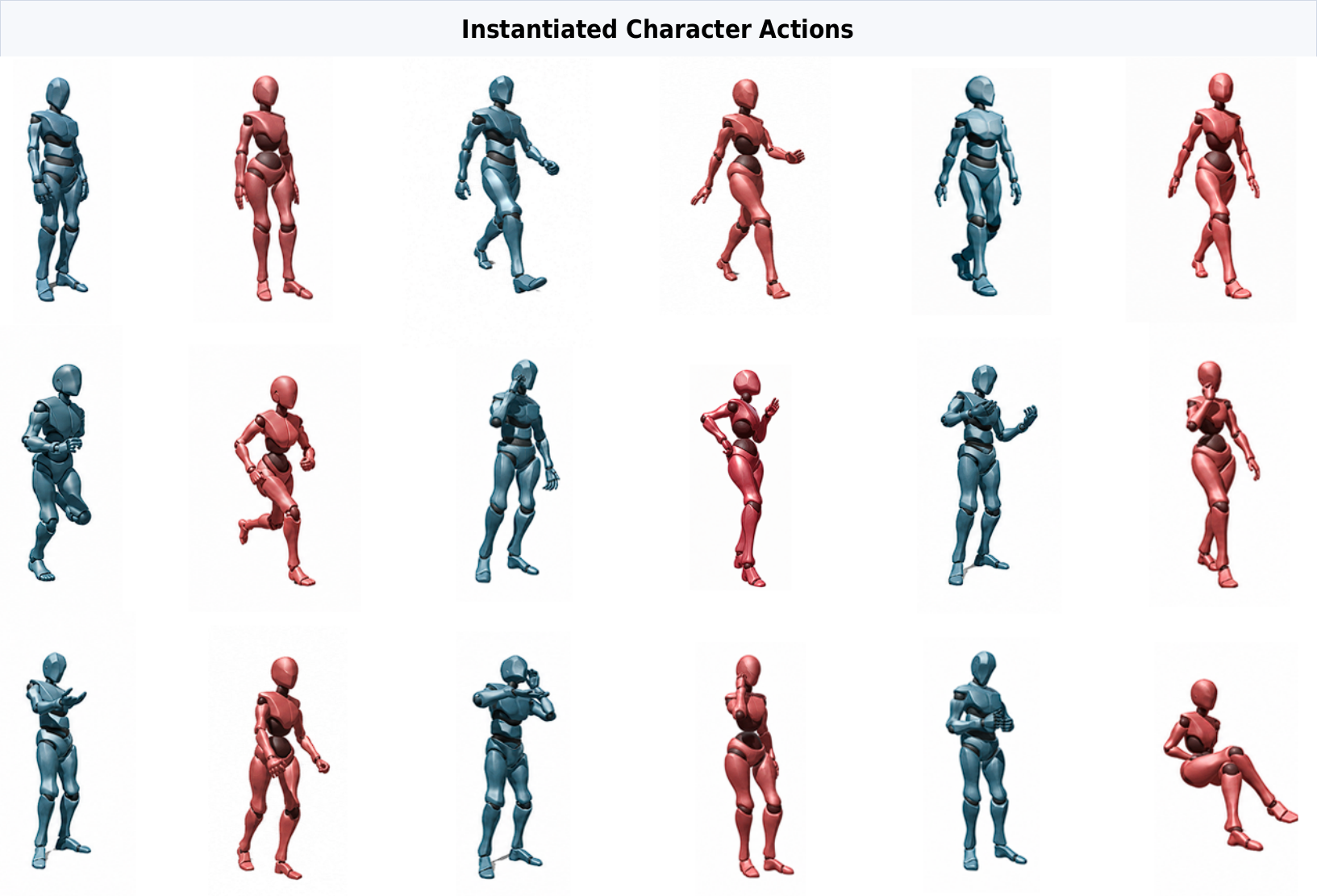}
  \caption{\textbf{Instantiated character actions.} Neutral mannequins display representative standing, walking, running, and gesturing configurations. Character resources and animation bindings are represented separately, with compatibility checked during integration.}
  \Description{A three-by-six grid of blue and red humanoid mannequins shows standing and locomotion configurations, expressive upper-body poses, gestures, and a seated pose.}
  \label{fig:character-actions}
\end{figure}

Figures~\ref{fig:character-fidelity} and~\ref{fig:character-actions} illustrate complementary resource dimensions. The character examples span high, medium, and low visual fidelity. The action examples show poses and motions on neutral mannequins. Keeping these dimensions distinct aligns with the entity and animation operations used in selective revision: appearance is associated with the selected character resource, whereas action is associated with an animation binding and its timing.

Character appearance and motion can be revised separately, subject to compatible resources and valid bindings. The demonstrated animation edits re-synchronize motion with existing transform tracks; unsupported asset combinations remain a potential source of failure.

\subsection{Geometry-Constrained Placement}
\label{subsec:spatial}

The spatial agent maps language such as ``on the roof'' or ``beside the statue'' to constraints over entity transforms and axis-aligned bounding boxes. Entity $i$ is represented by transform $\mathbf{T}_i=(\mathbf{p}_i,\mathbf{R}_i,\mathbf{s}_i)$, where $\mathbf{p}_i\in\mathbb{R}^3$ is position, $\mathbf{R}_i\in SO(3)$ is rotation, and $\mathbf{s}_i\in\mathbb{R}^3$ is scale. Its bounding box $\mathcal{B}_i=(\mathbf{p}_i,\mathbf{h}_i)$ uses half-extents $\mathbf{h}_i\in\mathbb{R}^3$. The agent translates semantic relations into relational constraints $\mathcal{C}_{\mathrm{rel}}$ and collision constraints $\mathcal{C}_{\mathrm{coll}}$. For example, ``$i$ on $j$'' induces vertical contact and horizontal-overlap constraints,
\begin{align}
[\mathbf{p}_i]_z-[\mathbf{h}_i]_z &= [\mathbf{p}_j]_z+[\mathbf{h}_j]_z,\nonumber\\
\big|[\mathbf{p}_i]_q-[\mathbf{p}_j]_q\big| &\le [\mathbf{h}_i]_q+[\mathbf{h}_j]_q,\qquad q\in\{x,y\},\nonumber
\end{align}
while ``$i$ beside $j$'' imposes proximity $\lVert[\mathbf{p}_i]_{xy}-[\mathbf{p}_j]_{xy}\rVert\le\delta$, with $\delta$ derived from entity extents; collision constraints additionally prevent geometric interpenetration.

When several transforms satisfy these constraints, the system selects a context-compatible placement by solving
\begin{equation}
\mathbf{T}_i^*=
\arg\min_{\mathbf{T}_i}
\Psi(\mathbf{T}_i;\mathcal{M},\mathcal{E})
\quad\text{subject to}\quad
\mathcal{C}_{\mathrm{rel}},\mathcal{C}_{\mathrm{coll}},
\label{eq:transform_optimization}
\end{equation}
where $\Psi$ is a placement objective informed by environmental context obtained through middleware $\mathcal{M}$ from engine $\mathcal{E}$, including support geometry and clearance from nearby boundaries. The resulting candidate is checked against the realized engine state. If validation fails, the spatial agent revises the constraints and re-solves the local placement problem. Algorithm~\ref{alg:spatial_placement} summarizes this loop.

\begin{algorithm}[t]

\caption{Geometry-Constrained Transformation with Engine Validation}
\label{alg:spatial_placement}
\begin{algorithmic}[1]
\REQUIRE entity specification $\mathcal{T}_{\mathrm{entity}}$, middleware $\mathcal{M}$, engine $\mathcal{E}$, spatial agent $\mathcal{A}_{\mathrm{spatial}}$
\ENSURE validated transforms $\{\mathbf{T}_i^*\}$ or a localized placement failure
\STATE obtain environmental context $\mathbf{C}_{\mathrm{env}}\leftarrow \mathcal{M}(\mathcal{E})$
\FOR{each entity $i$ in $\mathcal{T}_{\mathrm{entity}}$}
    \STATE construct $\mathcal{C}_{\mathrm{rel}}^i,\mathcal{C}_{\mathrm{coll}}^i\leftarrow \mathcal{A}_{\mathrm{spatial}}(\mathcal{T}_{\mathrm{entity}},\mathbf{h}_i,\mathbf{C}_{\mathrm{env}})$
    \STATE solve $\mathbf{T}_i$ using Eq.~\eqref{eq:transform_optimization}
    \WHILE{$\neg\operatorname{validate}_{\mathcal{E}}(\mathbf{T}_i)$ and the local refinement budget remains}
        \STATE relax or revise the violated constraints
        \STATE re-solve $\mathbf{T}_i$ using updated $\mathcal{C}_{\mathrm{rel}}^i,\mathcal{C}_{\mathrm{coll}}^i$
    \ENDWHILE
    \IF{$\operatorname{validate}_{\mathcal{E}}(\mathbf{T}_i)$}
        \STATE store $\mathbf{T}_i^*\leftarrow\mathbf{T}_i$
    \ELSE
        \STATE return a localized placement failure for entity $i$
    \ENDIF
\ENDFOR
\RETURN $\{\mathbf{T}_i^*\}$
\end{algorithmic}
\end{algorithm}

This procedure keeps placement reasoning local to the affected entities while using engine validation to reject transforms that are inconsistent with the realized scene. The constraints model geometry and support relations; they do not constitute a general rigid-body physics solver.

\subsection{Capability-Aligned Code Synthesis}

The knowledge library contains more than 60 verified code templates covering world editing, sequence initialization, track management, keyframes, camera control, animation, and synchronization. Each entry includes a descriptor, typed parameters, a validated code body, and compatibility or execution metadata. Middleware dispatches an agent-specific view: camera agents receive camera patterns, animation agents receive animation patterns, and sub-agents inherit common Sequencer operations.

For a subtask $v$, synthesis is conditioned on the relevant template view and execution context:
\begin{equation}
\mathtt{code}_v = \Pi_{\mathcal{A}}\!\left(\mathcal{K}^{\mathcal{A}}_v, v \mid \mathbf{C}^{\mathcal{A}}_v\right).
\label{eq:code_synthesis}
\end{equation}
Here, $\mathcal{K}^{\mathcal{A}}_v$ is the retrieved template subset and $\mathbf{C}^{\mathcal{A}}_v$ contains Sequencer state, execution scope, and required constraints, including synchronization dependencies. The language model selects and instantiates relevant patterns rather than treating the entire engine API as an unconstrained code-generation space.

Assembly proceeds at three levels. The Sequencer agent creates the common timeline and track structure. Specialized sub-agents generate self-contained functions for camera trajectories, transform animation, and animation playback. Template instantiation produces parameterized calls with exception handling and checks against applicable engine interfaces. Middleware coordinates the assembled functions and supported transactional execution. Capability isolation limits irrelevant patterns, while shared context prevents isolation from becoming a loss of cross-track information.

\subsection{Camera and Multi-Shot Coordination}

A camera operation combines position and orientation with focal length, aperture, and focus distance. The camera agent maps the directive to keyframes and interpolation within Sequencer. A dolly zoom illustrates why the relationship between controls matters: maintaining approximately constant projected subject size requires coordinated camera displacement and focal-length changes,
\begin{align}
\mathbf{p}_{\mathrm{cam}}(t) &= \mathbf{p}_{\mathrm{cam}}(t_0)+(d(t)-d_0)\mathbf{n},\nonumber\\
f(t) &= f_0\frac{d(t)}{d_0},
\label{eq:dolly}
\end{align}
where $d(t)$ is camera--subject distance, $d_0$ and $f_0$ are the initial distance and focal length, and $\mathbf{n}$ is the unit direction from the subject toward the initial camera position (the direction of increasing $d$). The relation uses a fixed-subject, perspective-projection approximation and supplies coupled position and focal-length values for the camera tracks.

The camera agent identifies the coupling, computes aligned position and focal-length keyframes, and generates transform and lens tracks. Multi-shot directions are represented by temporally bounded camera actors and cut-track segments. Shared timeline context coordinates these with actor animations. Framing heuristics and focus tracking provide compositional defaults, while the resulting camera actors, lens values, and cut boundaries remain available for inspection and revision.

\subsection{Verification-Grounded Repair}
\label{subsec:verification}

Verification operates at three stages. \emph{Static validation} checks syntax, types, and API compatibility before execution. \emph{Instrumented execution} records calls, parameters, returns, and exceptions. \emph{Post-execution state inspection} compares the realized scene, transforms, tracks, and camera parameters with task expectations. These stages locate failures before, during, and after execution. Section~\ref{sec:metrics} separately defines the outcomes used to assess task completion.

On a failed check, middleware classifies the diagnostic information and invokes the responsible agent for function-level revision. Validated operations are retained where the failing function can be isolated. Iteration is bounded, and the system supports simplified fallbacks and diagnostic reporting when refinement stalls. Two-version reflection rollback provides a recovery path for revised operations. Rollback operates on the revised functions and their associated state supported by the middleware.

We formalize the verification-reflection update as
\begin{equation}
\Gamma_{i+1}=\mathcal{M}\!\left(
\mathcal{E}\!\left(
\Pi_{\mathcal{A}^*}(\mathcal{K},v\mid\mathbf{C}_v,\Gamma_i)
\right)\right),
\label{eq:verification_reflection}
\end{equation}
where $\Gamma_i$ is the verification state at refinement step $i$, $\mathcal{A}^*$ is the specialized agent responsible for the failed operation, $\mathcal{K}$ is the verified knowledge library, and $\mathbf{C}_v$ is the task and execution context. The responsible agent regenerates the affected function conditioned on the diagnostic state; $\mathcal{E}$ realizes the revised operation and $\mathcal{M}$ evaluates the next verification state $\Gamma_{i+1}$. The loop terminates when the checks pass or the bounded refinement procedure returns a fallback or diagnostic failure.

\paragraph{Evidence and decision boundaries.}
The update in Equation~\eqref{eq:verification_reflection} is useful only when diagnostic feedback distinguishes the kind of failure that occurred. Table~\ref{tab:verification-decisions} makes that distinction explicit. A call can be well formed yet fail at runtime, and a completed call can leave a binding or temporal relation inconsistent with the request. Aesthetic judgment addresses a different question: whether a technically valid result serves the author's composition.

\begin{table}[!htbp]
\caption{\textbf{Evidence and verification decisions.} Static checks, runtime traces, and world-state inspection support different repair decisions. Visual inspection evaluates composition and may trigger a revised specification; it is distinct from automated verification.}
\Description{Four rows connect a decision boundary to the evidence examined, the question answered, and the appropriate response. Static checks use generated code and API signatures. Runtime diagnosis uses calls, returns, and exceptions. State inspection uses actors, bindings, geometry, and timing. Author inspection uses the viewport and timeline to judge the composition and retain or revise the brief.}
\label{tab:verification-decisions}
\begingroup
\small
\renewcommand{\arraystretch}{1.16}
\setlength{\tabcolsep}{5pt}
\begin{tabularx}{\linewidth}{>{\raggedright\arraybackslash}p{0.17\linewidth}YYY}
\toprule
\rowcolor{tablehead}\textbf{Decision boundary} & \textbf{Evidence examined} & \textbf{Question addressed} & \textbf{Resulting response} \\
\midrule
\textbf{Static validity} & Generated syntax, argument types, and supported API signatures. & Is the proposed operation well formed for the available interface? & Revise the failing function before submitting it to the engine. \\
\addlinespace[4pt]
\textbf{Runtime completion} & Instrumented calls, parameters, return values, and exceptions. & Did the submitted operation complete, and where did execution fail? & Localize the failing call and provide its trace to the responsible agent. \\
\addlinespace[4pt]
\textbf{State conformance} & Realized actors, asset bindings, transforms, keyframes, and track timing. & Does the resulting state satisfy the represented spatial and temporal requirements? & Revise the responsible operation using engine-state feedback; retain validated operations. \\
\midrule
\rowcolor{tablenote}\textbf{Creative judgment} & The rendered viewport and inspectable scene and timeline. & Does the realized composition serve the author's intended expression? & Retain the result or revise the brief through language or native engine editing. \\
\bottomrule
\end{tabularx}
\endgroup
\end{table}

A useful diagnostic therefore connects \emph{evidence}, an \emph{affected operation}, and a \emph{world or timeline object}. Function-level repair uses this connection to revisit a concrete execution unit. A subsequent user instruction can target the same unit but modify the requirement itself. Diagnostic repair seeks conformance to an existing specification, whereas a new instruction changes that specification.

\FloatBarrier
\section{Technical Evaluation}
\label{sec:evaluation}

We evaluate scene construction, selective revision, temporal coordination, and the contributions of specialized knowledge, verification, and localized repair. Section~\ref{sec:discussion} examines the resulting system properties and limitations.

\subsection{Implementation and Evaluation Measures}
\label{sec:metrics}

\cineu uses Claude Sonnet and Gemini as agent backbones~\cite{anthropic2025claude37sonnet,comanici2025gemini}. The Unreal Engine backend supports version 5.5 and earlier versions through programmatic interfaces and Blueprint integration. The resource layer contains more than 300 validated assets and more than 60 verified code templates. The illustrated tasks combine environment configuration, entity placement, animation, camera movement, and shot timing. The single-agent baseline removes hierarchical coordination and specialized sub-agents; the remaining comparisons examine knowledge access, verification, and repair within the multi-agent system.

\paragraph{Task success.}
End-to-end completion requires three outcomes: execution validity, including executor and communication behavior; correct assets and animation attributes; and satisfaction of spatial and temporal constraints, including animation--transform synchronization. Middleware and engine interfaces provide the checks. These outcome criteria are broader than whether an initial program is syntactically executable.

\paragraph{Program correctness and refinement.}
Program correctness measures the correctness of the initial generation before iterative refinement. Refinement cycles describe the iterative corrections associated with each configuration. Reporting these alongside task success separates first-attempt synthesis from the recovery process that follows it.

\paragraph{Comparison structure.}
Table~\ref{tab:ablation} presents configuration-level results, which we analyze descriptively. The knowledge variants use no curated knowledge, a unified library, or capability-aligned subsets. The verification variants use no verification, a single stage, or the static--execution--state chain. The repair variants regenerate the full program or revise the failing function. The qualitative analysis follows the corresponding scene and timeline operations through the displayed examples, with the video comparison focusing on camera--lens coupling and scene continuity.

\subsection{Scene Construction and Multi-Shot Execution}
\label{sec:expressiveness}

Figure~\ref{fig:overview} presents three compound briefs: an architectural reveal, a rooftop sequence with coupled lens motion and an orbit, and a change of attention between objects above and below water. Each brief combines content with camera behavior, rather than only specifying a desired image.

\begin{figure}[!htbp]
  \centering
  \includegraphics[width=0.78\linewidth]{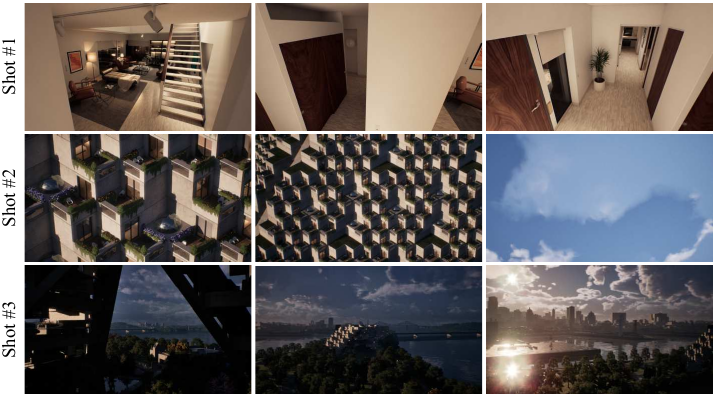}
  \caption{\textbf{Multi-shot coordination.} Three camera shots expand the scene from interior exploration to an exterior panorama. Camera actors and cut-track segments are coordinated through a shared timeline rather than assembled as independently generated clips.}
  \Description{Three rows labeled Shot 1, Shot 2, and Shot 3 show successive camera views of an interior, an exterior building arrangement, and a city skyline.}
  \label{fig:multishot}
\end{figure}

Figure~\ref{fig:multishot} shows a three-shot sequence that expands from interior exploration through exterior building views to a cityscape panorama. Distinct camera trajectories and cut-track segments share temporal context. The resulting camera actors and shot boundaries remain represented in Sequencer, connecting the displayed frames to an inspectable temporal structure.

The available character resources in Figures~\ref{fig:character-fidelity} and~\ref{fig:character-actions} show variation in available appearance and motion. The level-simulation examples in Figure~\ref{fig:orchestration}(b) further demonstrate locomotion and pose transitions, including walking and jumping. Together, the examples show how entity resources, action bindings, and simulation interfaces support character behavior in the same engine environment used for cinematic authoring.

\subsection{State-Preserving Revision}
\label{sec:selective-revision}

\begin{figure}[!htbp]
  \centering
  \includegraphics[width=0.76\linewidth]{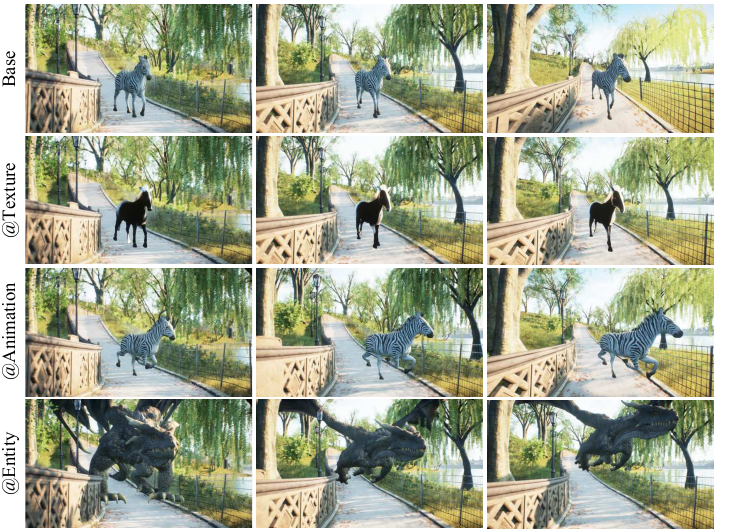}
  \caption{\textbf{Selective revision of a shared scene.} The base sequence is modified along texture, animation, and entity dimensions. Animation changes require synchronization with displacement; entity replacement requires placement checks. The environment and broader camera composition are retained in the displayed examples.}
  \Description{A four-row frame grid shows a base zebra sequence followed by texture, animation, and entity substitutions. The rows retain corresponding environment and camera views while changing the selected scene component.}
  \label{fig:editing}
\end{figure}

The revisions in Figure~\ref{fig:editing} share a base scene. Texture substitution changes surface appearance without reconstructing the temporal sequence. Animation replacement changes the motion clip and aligns it with existing transform motion. Entity replacement substitutes an asset and re-establishes its spatial relationship to the environment. Corresponding frames retain the broader environment and camera viewpoints while exposing the intended change.

These cases show that the scope of an update is determined by its dependencies. Texture edits have a narrower operational scope than entity edits: the latter may change geometry and therefore require placement checks. Animation replacement must account for temporal coupling even when entity identity remains fixed. In these cases, retaining the surrounding composition requires changes to the targeted resource together with any directly coupled placement or timing operations. Selective revision is therefore realized through the relation between a requested change and its engine dependencies.

\subsection{Camera Control and Cross-Track Coordination}
\label{sec:coordination}

\begin{figure}[!htbp]
  \centering
  \includegraphics[width=\linewidth]{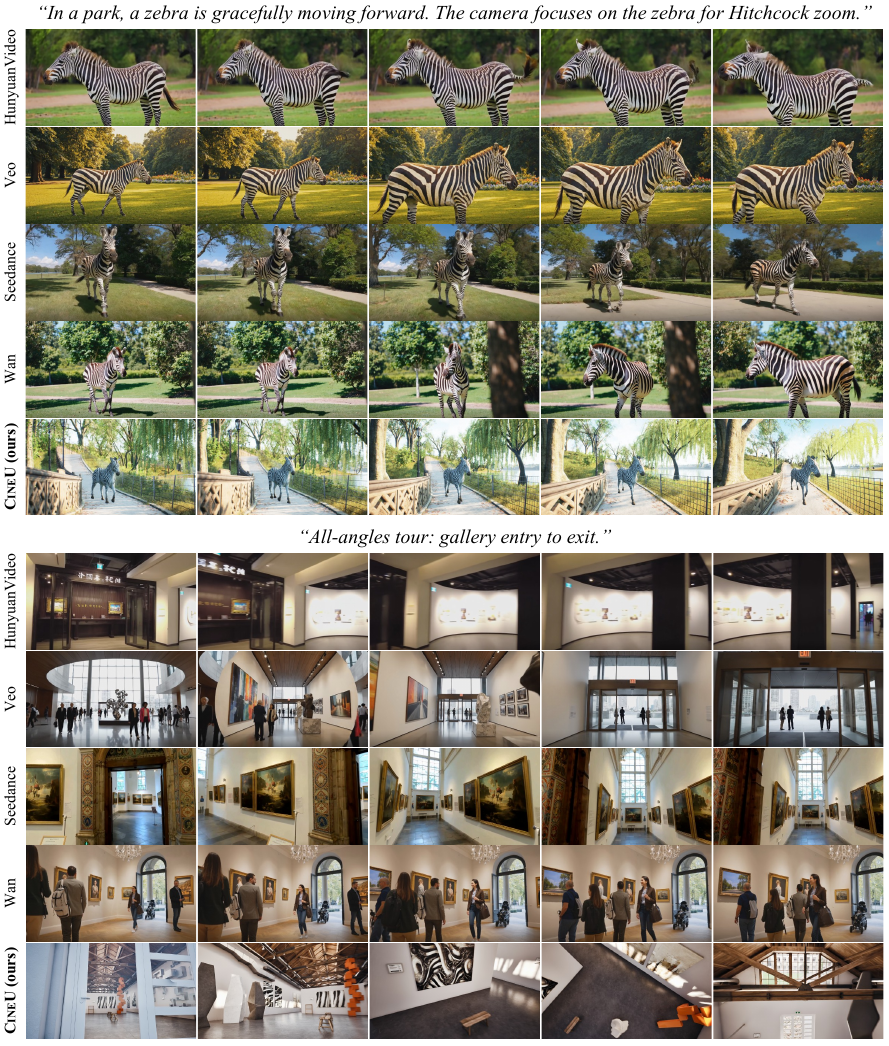}
  \caption{\textbf{Illustrative camera-control cases.} Outputs for a zebra dolly-zoom request and a continuous gallery tour compare generative video with engine-grounded authoring. The selected examples emphasize camera--lens coupling and scene continuity, contrasting generated frames with an explicit engine scene and camera trajectory.}
  \Description{Two frame grids compare HunyuanVideo, Veo, Seedance, Wan, and CineU. The first grid shows zebra scenes under a dolly-zoom prompt; the second shows gallery views under an entrance-to-exit tour prompt.}
  \label{fig:comparison}
\end{figure}

Figure~\ref{fig:comparison} contrasts the example outputs of HunyuanVideo, Veo, Seedance, Wan, and \cineu for two prompts. In the zebra example, the target effect is a dolly zoom. The generative outputs provide richer appearance in the selected frames, while \cineu explicitly couples distance and focal length through Equation~\ref{eq:dolly}. In the gallery example, \cineu executes a trajectory through one persistent scene, retaining its camera waypoints and parameters. The comparison highlights complementary representations: generative models supply appearance through learned priors, while \cineu retains scene identity and camera parameters in an editable engine sequence.

\begin{figure}[!htbp]
  \centering
  \begin{minipage}[t]{0.49\linewidth}
    \centering
    \includegraphics[width=\linewidth]{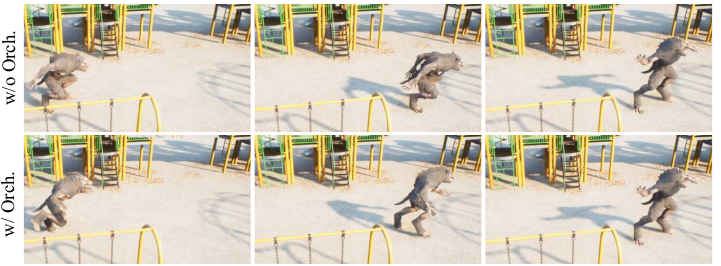}\par
    \small (a) Isolated and coordinated tracks
  \end{minipage}\hfill
  \begin{minipage}[t]{0.49\linewidth}
    \centering
    \includegraphics[width=\linewidth]{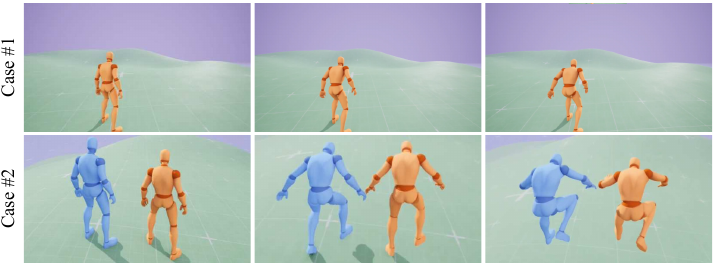}\par
    \small (b) Level-simulation character control
  \end{minipage}
  \caption{\textbf{Coordination and character control.} (a) Without orchestration (upper row), the wolf translates with a static pose; coordinated tracks (lower row) align animation with movement. (b) Two level-simulation examples demonstrate locomotion and pose transitions. Both panels are qualitative system examples.}
  \Description{At left, two rows of wolf frames contrast static-pose sliding with coordinated locomotion. At right, two rows labeled Case 1 and Case 2 show humanoid characters moving and changing pose in a simple simulated environment.}
  \label{fig:orchestration}
\end{figure}

Figure~\ref{fig:orchestration}(a) isolates a temporal dependency. Without Sequencer-level coordination, transform and animation agents produce spatial displacement and a static pose, causing the wolf to slide. With shared timing context, locomotion progresses with displacement and transitions to the next animation state. The relevant distinction is not whether either track exists, but whether their relationship realizes the action. Specialized knowledge and shared context address different problems: the former constrains the operations available to an agent, while the latter connects operations that must agree.

\subsection{Component Ablations and Recoverability}

Table~\ref{tab:ablation} reports the configuration comparisons. The vanilla single agent reaches $11.5\%$ task success and $19.4\%$ first-attempt program correctness, with the iteration entry recorded as $>10$. Among the knowledge configurations, success is $18.4\%$ without knowledge, $31.2\%$ with a unified library, and $47.8\%$ with specialized knowledge. The latter is approximately $2.6\times$ the no-knowledge value. The ordering is consistent with the design of capability-aligned retrieval: verified patterns provide executable examples, and specialization reduces irrelevant patterns within each agent's context.

\begin{table}[!htbp]
\centering
\caption{\textbf{Component ablations.} Success is end-to-end task completion, program correctness concerns the first generation, and refinement cycles describe iterative correction. Configuration-level values are organized into knowledge, verification, and repair comparisons. Bold entries identify the full-system configuration within each group.}
\Description{Nine configurations are compared on success, program correctness, and refinement cycles. The specialized-knowledge, multi-stage-verification, and minimal-intervention rows each report 47.8 percent, 56.2 percent, and 5.6 cycles. These rows repeat the full configuration in different comparison groups.}
\label{tab:ablation}
\begin{tabular}{lccc}
\toprule
\textbf{Configuration} & \textbf{Success (\%)} & \textbf{Program (\%)} & \textbf{Refinement Cycle} \\
\midrule
\textcolor{basecolor}{Vanilla Agent} & \textcolor{basecolor}{11.5} & \textcolor{basecolor}{19.4} & \textcolor{basecolor}{$>$10}\\
w/o Knowledge & 18.4 & 22.6 & 8.7 \\
w/ Unified $\mathcal{K}$ & 31.2 & 38.5 & 7.3 \\
w/ \textit{Specialized} $\mathcal{K}^{\mathcal{A}}_v$ & \textbf{47.8} & \textbf{56.2} & \textbf{5.6} \\
\grayrule
w/o Verification & 16.8 & 56.2 & 9.4 \\
w/ Single-Stage & 29.5 & 56.2 & 6.9 \\
w/ \textit{Multi-Stage} & \textbf{47.8} & \textbf{56.2} & \textbf{5.6} \\
\grayrule
Full Regeneration & 38.1 & 56.2 & 7.8 \\
\textit{Minimal Intervention} & \textbf{47.8} & \textbf{56.2} & \textbf{5.6} \\
\bottomrule
\end{tabular}
\end{table}

Verification affects recovery rather than the recorded first attempt: program correctness remains $56.2\%$ across the verification rows, while success changes from $16.8\%$ without verification to $29.5\%$ with one stage and $47.8\%$ with multiple stages. The corresponding refinement entries are $9.4$, $6.9$, and $5.6$. Syntax checks address errors that can be detected before execution; tracing and state inspection additionally expose failures in realized behavior.

Minimal-intervention repair records $47.8\%$ success compared with $38.1\%$ for full regeneration, an absolute difference of $9.7$ percentage points. Its refinement entry is $5.6$ rather than $7.8$. This pattern is consistent with retaining valid operations during repair: the revision remains focused on the diagnosed function rather than reopening the complete program. The remaining failures motivate the analysis of invalid placement, framing, and cross-track interference below.

\subsection{Failure Analysis}

\begin{figure}[H]
  \centering
  \includegraphics[width=0.94\linewidth]{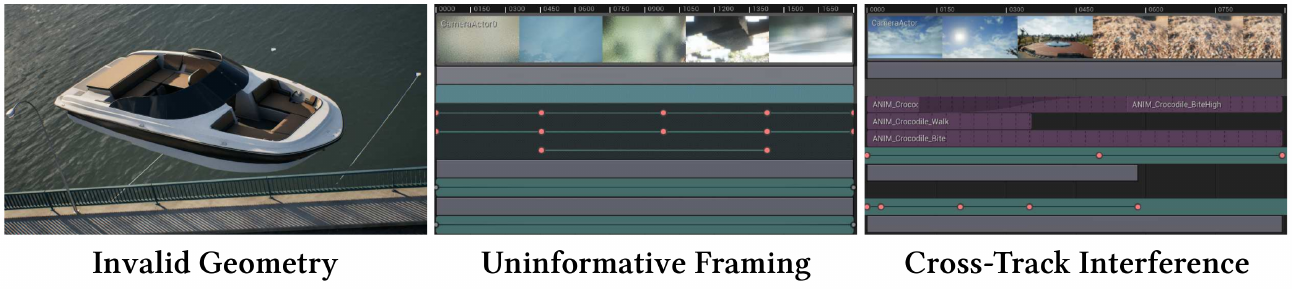}
  \caption{\textbf{Representative failures.} Invalid placement, uninformative camera framing, and interference between tracks illustrate different gaps between executable operations and the intended scene. Each panel connects a visible scene or timeline issue to the operation requiring correction.}
  \Description{Three panels show a boat placed incorrectly near a bridge, a camera timeline associated with uninformative framing, and interfering or desynchronized tracks in Unreal Sequencer.}
  \label{fig:failures}
\end{figure}

Figure~\ref{fig:failures} shows invalid geometry, uninformative framing, and cross-track interference. The baseline analysis additionally identifies contextual mixing between operation domains and oscillation during repair. A camera section may execute correctly yet frame an irrelevant part of the scene; a valid transform and a valid animation can still produce an incoherent event. These cases motivate checking relationships as well as individual operations.

Asset gaps, incompatible resources, and unresolved geometric constraints can also prevent completion. A diagnostic fallback may preserve a usable partial scene, but a relaxed constraint is not equivalent to satisfying the original request. The diagnostic status should distinguish operational failure, constraint relaxation, and differences in visual preference rather than treating them as the same outcome.

\FloatBarrier
\section{Discussion}
\label{sec:discussion}

The results illustrate how agentic reasoning can be grounded in a simulator that exposes structured state. The relevant technical properties are the addressability of generated objects, the dependencies between operations, and the reliability of execution feedback. These properties determine whether a system can modify an existing world without losing previously validated work.

\subsection{Persistent State as an Execution Interface}

Unlike a sequence of generated frames, an Unreal project retains explicit actors, transforms, animation bindings, camera parameters, and temporal tracks. These objects provide targets for subsequent operations and diagnostics. A material substitution can address a particular actor; a camera revision can address a specific track and interval. The rendered viewport remains useful for inspecting appearance, but it does not replace the engine state needed to identify and modify the operation that produced a result.

This distinction connects to work on editable intermediate representations in AI Chains, VRCopilot, and Dreamcrafter~\cite{wu2022aichains,zhang2024vrcopilot,vachha2024dreamcrafter}. In \cineu, the intermediate representation is executable scene and timeline state rather than only a visible plan or proxy. The resulting benefit is not that every requested edit is independent. It is that the system can identify the objects, constraints, and functions implicated by the edit.

\subsection{Update Locality Depends on Relationships}

The selective-revision cases show that the scope of a modification cannot be defined solely by the resource named in the instruction. A replacement entity may have different bounds and require a placement adjustment to preserve support contact. A new animation must remain synchronized with the actor's transform track. In contrast, a texture update can preserve the existing motion and camera sequence. Table~\ref{tab:revision-dependencies} makes these relationships explicit.

Locality is therefore \emph{relational}: an update should preserve validated constraints while changing the smallest set of operations needed to satisfy the revised instruction. This is also the rationale for combining capability-specific agents with shared context. Specialized knowledge limits the API patterns considered by each agent; shared object and timeline references maintain consistency across their outputs. The two mechanisms address different failure modes and should not be conflated.

\subsection{Verification, Recovery, and Their Boundaries}

A syntactically valid program, a completed engine call, and a correct scene are different outcomes. Static analysis checks the proposed operation; tracing reveals runtime behavior; and engine-state inspection evaluates observable scene and timing constraints. Table~\ref{tab:verification-decisions} describes the evidence available at each stage. The ablations indicate that the verification chain affects recovery even when first-attempt program correctness is unchanged across configurations.

Function-level repair retains validated operations when the failure can be isolated. This reduces the scope of regeneration, but its effectiveness depends on whether the diagnostic identifies the responsible function and affected state. Cross-track errors may require coordinated updates, and an invalid asset or unresolved geometric constraint may not admit a local fix. Moreover, state conformance is not equivalent to visual quality: a camera can satisfy formal parameter constraints while producing an undesirable composition. The system checks operational requirements; choosing or revising an artistic objective remains a separate decision.

\subsection{Relation to Code-Based World Models}

Executable world representations offer a way to make state changes inspectable and repeatable. \cineu adopts this perspective at the control layer: it produces code that modifies a persistent world, but the world state and dynamics are provided by Unreal rather than inferred from observations. This differs from learning a transition model or synthesizing a new simulator. The distinction also defines a verification boundary. Successful code execution does not establish that an agent has learned physical laws; it establishes only that the specified operations can be checked against available engine state and task constraints. In this setting, the benefit of code is to expose editable objects, transitions, and their dependencies for continued orchestration.

\subsection{Limitations and Scope}

The system depends on the availability and compatibility of its assets, templates, and Unreal interfaces. Axis-aligned bounding boxes approximate support and separation but do not constitute a general contact or rigid-body physics solver. Camera geometry uses simplified assumptions, and the reported configuration-level ablations do not establish universal performance across all prompts, engines, or asset libraries. Reproduction depends on committed project state, compatible engine versions, and deterministic components under fixed settings. Two-version rollback applies only to the functions and state covered by the middleware.

Engine-grounded generation complements, rather than replaces, learned visual synthesis. Generated video can offer richer appearance, while structured simulation supports explicit geometry, timing, and localized revisions. Combining these strengths is a useful direction for future systems, but would require additional methods for transferring generated content into compatible assets and executable state. Asset provenance, licensing, and potentially deceptive media uses also remain considerations for deployment.

\section{Conclusion}

We introduced \cineu, an agentic system that translates language instructions into executable operations over persistent Unreal Engine worlds. Its hierarchical orchestration coordinates asset retrieval, geometry-constrained placement, animation, camera trajectories, and shared timelines. Capability-aligned code synthesis grounds operations in verified engine patterns, while staged verification and function-level repair use execution feedback to recover from failures without unnecessarily regenerating valid state. Scene construction, multi-shot examples, selective revisions, and component ablations characterize the system's controllability and recovery behavior. Viewed alongside code-based world modeling, \cineu demonstrates an engine-grounded approach in which agent-generated programs control a persistent world representation rather than replace the simulator's dynamics. Precise world simulation requires more than independent valid actions: it requires explicit dependencies and retained scene state so that operations remain consistent across execution and revision.

\renewcommand{\bibfont}{\small}
\bibliographystyle{unsrtnat}
\bibliography{references}
\end{document}